\documentclass[%
 reprint,twocolumn,
superscriptaddress,
nobibnotes,
 amsmath,amssymb,
 aps,
 prl,
]{revtex4-2}

\usepackage[mathscr]{euscript}
\usepackage{color}
\usepackage{graphicx}
\usepackage{dcolumn}
\usepackage{bm}
\usepackage{appendix}

\begin{document}

\title{Higher-order correlations reveal complex memory in temporal hypergraphs\\
}

\author{Luca Gallo}
\email{gallol@ceu.edu}
\affiliation{%
 Department of Network and Data Science, Central European University, Vienna, Austria
}%
\author{Lucas Lacasa}
\affiliation{%
Institute for Cross-Disciplinary Physics and Complex Systems (IFISC), CSIC-UIB, Palma de Mallorca, Spain
}%
\author{Vito Latora}
\affiliation{School of Mathematical Sciences, Queen Mary University of London, London E1 4NS, UK}
\affiliation{Department of Physics and Astronomy,  University of Catania, 95125 Catania, Italy}
\affiliation{INFN Sezione di Catania, Via S. Sofia, 64, 95125 Catania, Italy}
\affiliation{Complexity Science Hub Vienna, A-1080 Vienna, Austria}
\author{Federico Battiston}
\email{battistonf@ceu.edu}%
\affiliation{%
 Department of Network and Data Science, Central European University, Vienna, Austria
}%

\date{\today}

\begin{abstract}
Many real-world complex systems are characterized by interactions in groups that change in time. 
Current temporal network approaches, however, are unable to describe group dynamics, as they are based on pairwise interactions only.
Here, we use time-varying hypergraphs to describe such systems,  and we introduce a framework based on higher-order correlations to characterize their temporal organization.
We analyze various social systems, finding that groups of different sizes have typical patterns of long-range temporal correlations. Moreover, our method reveals the presence of non-trivial temporal interdependencies between different group sizes. 
We introduce a model of temporal hypergraphs with non-Markovian group interactions, which reveals complex memory as a fundamental mechanism underlying the pattern in the data.
\end{abstract}

\maketitle

Temporal networks, where links connecting pairs of nodes are not continuously active, provide a framework to model how the interactions of a complex system evolve in time~\cite{holme2012temporal, masuda2016guide, holme2019temporal}.  
They have revealed key in understanding how the time-varying interaction network of real-world social and biological systems affects the properties of dynamical processes, such as epidemic spreading~\cite{karsai2011small, van2013non, rosvall2014memory, masuda2017introduction}, diffusion~\cite{perra2012random, starnini2012random, masuda2013temporal, scholtes2014causality, delvenne2015diffusion}, synchronization~\cite{kohar2014synchronization,zhang2021designing}, and others~\cite{zanin2009dynamics, starnini2013modeling, fallani2008persistent, peixoto2017modelling}. 
Recent results have highlighted the complex way in which the activity of each link depends on the activities of all other links, showing that memory~\cite{vestergaard2014memory,p2018change,williams2019effects,williams2022non} in temporal networks is inherently a multidimensional concept with a well defined microscopic shape~\cite{williams2022shape}. 
Different approaches have aimed to describe the time evolution of a network as a trajectory in graph space, by naturally extending to the case of graphs notions such as correlations~\cite{lacasa2022correlations} or even dynamical stability~\cite{caligiuri2023lyapunov} traditionally used for scalar or vectorial time-series.

Temporal network approaches, however, have a strong limitation. They are based on a network description and, as such, they can only describe how dyadic interactions (i.e., links) vary in time. 
On the other hand, many real-world  social~\cite{benson2018simplicial, patania2017shape, cencetti2021temporal,lotito2022higher,contisciani2022inference}, biological~\cite{klamt2009hypergraphs,zimmer2016prediction}, neural~\cite{petri2014homological,giusti2016two} and ecological~\cite{levine2017beyond, grilli2017higher} systems exhibit higher-order interactions, i.e., interactions involving groups of three or more units at the same time. 
Such many-body interactions are better modeled by higher-order networks, such as hypergraphs and simplicial complexes, where hyperedges and simplices encode interactions among an arbitrary number of units~\cite{battiston2020networks, battiston2021physics}. 
Interestingly, taking into account the higher-order architecture of real-world systems is known to produce novel collective phenomena in a variety of dynamical processes, including diffusion~\cite{schaub2020random,carletti2020random}, synchronization~\cite{skardal2019abrupt, millan2020explosive, lucas2020multiorder,gambuzza2021stability,carletti2020dynamical}, contagion~\cite{iacopini2019simplicial,st2021universal} and evolutionary games~\cite{alvarez2021evolutionary,civilini2021evolutionary}.

Some early works have already started to explore the temporal dimension of higher-order interactions. For instance, group interactions in real-world social systems have been found to occur in persistent bursts of activity~\cite{cencetti2021temporal}, with events of different sizes close in time also spatially correlated in the network~\cite{ceria2022temporal}. 
Such persistent temporal higher-order interactions have been shown to anticipate the onset of endemic states in epidemic processes~\cite{chowdhary2021simplicial}, and to affect the convergence time of nonlinear consensus dynamics~\cite{neuhauser2021consensus}. 
Theoretical frameworks  
for modeling temporal group activation data~\cite{petri2018simplicial}, and for constructing simplicial complexes based on topological data analysis of multivariate time-series from brain functional activity, financial markets and disease spreading~\cite{santoro2023higher} have also been recently developed. 
However, how to analyze and characterize the temporal organization of real-world complex systems with higher-order interactions is to this day still an open problem.

\bigskip
In this Letter, we bridge this gap by introducing a general framework to study higher-order temporal dependencies in complex systems.  
We represent a complex system with interactions in groups whose size and composition can change in time as a temporal hypergraph, i.e., a hypergraph with time-varying hyperedges of different orders. 
We then define a set of measures to extract higher-order temporal correlations, namely to characterize how the dynamics of hyperedges of different orders are correlated. 
By focusing on a variety of empirical social systems, our framework reveals the existence of long-range correlations at different group sizes, and their hierarchical organization.
Furthermore, we uncover the presence of temporal correlations between groups of different sizes, i.e., between hyperedges of different orders, capturing emergent interactions between coherent mesoscopic structures. 
Finally, to gain intuition about the underlying microscopic mechanisms, we introduce novel models of temporal hypergraphs with higher-order memory, able to explain the observed empirical patterns. 

\vspace{3mm}
\noindent \emph{Temporal correlations in hypergraphs}. --- Systems 
with higher-order interactions can be represented by hypergraphs~\cite{berge1973graphs}. A hypergraph is a tuple $(\mathcal{V},\mathcal{H})$, where $\mathcal{V}$ is a set of $N$ nodes, and $\mathcal{H}$ is a set of $M$ hyperedges. Each hyperedge is a collection of nodes representing an interaction among multiple units. A hyperedge of order 2, or $2$-hyperedge, is a set of two nodes  representing a two-body interaction, a $3$-hyperedge is a set of three nodes representing a group interaction among three units, and so on, up to a order $D$. 
A hypergraph can be encoded using a set of adjacency matrices $A^{(2)}$, $A^{(3)}$, $\dots$, $A^{(D)}$,
where the element $a^{(d)}_{ij}$ of the matrix $A^{(d)}$ counts the number of $d$-hyperedges both nodes $i$ and $j$ belong to, while elements $a^{(d)}_{ii}$ are set to zero.
To describe the temporal evolution of higher-order interactions, we consider temporal hypergraphs~\cite{cencetti2021temporal}. 
A temporal hypergraph 
is an ordered sequence of (static) hypergraphs $(\mathcal{V},\mathcal{H}(t))$, with $t\in\{1,\dots,T\}$, where $\mathcal{H}(t)$ is the set of hyperedges existing at time $t$.
For each order $d$, we can define a sequence of adjacency matrices $\{A^{(d)}(t)\}_{t=1}^{T} = \{A^{(d)}(1), A^{(d)}(2),\dots, A^{(d)}(T)\}$ that represents the temporal evolution of the system interactions of order $d$.

The presence of higher-order interactions makes the analysis of temporal correlations a multi-faceted problem. 
First, to quantify temporal correlations in interactions of a given order $d$, we introduce the intra-order correlation matrix
\begin{equation}\label{eq:intra_order_correlation_matrix}
\footnotesize
\mathscr{C}^{(d)}(\tau) = \frac{1}{T-\tau}\sum\limits_{t=1}^{T-\tau}\frac{1}{(d-1)!^2}\left[A^{(d)}(t)-\mu^{(d)}\right]\cdot\left[A^{(d)}(t+\tau)-\mu^{(d)}\right]^\intercal,
\end{equation}
with $d\in\{2,\dots,D\}$. Here, $\tau$ is the temporal lag, $A^\intercal$ denotes the transpose of $A$, and we have defined the annealed adjacency matrix of order $d$ as $\mu^{(d)} = \frac{1}{T}\sum_{t=1}^{T}A^{(d)}(t)$. Note that, for $d=2$, Eq.~(\ref{eq:intra_order_correlation_matrix}) recovers the correlation matrix for temporal networks \cite{lacasa2022correlations}. 
The diagonal terms of $\mathscr{C}^{(d)}(\tau)$ capture how hyperedges of order $d$ are temporally autocorrelated, whereas the off-diagonal terms quantify cross-correlations. When the latter are negligible, one can focus on the diagonal terms and define an intra-order correlation function
\begin{equation}\label{eq:intra_order_correlation_function}
c^{(d)}(\tau) = \mathrm{tr}(\mathscr{C}^{(d)}(\tau)),   
\end{equation}
that provides a scalar measure of how hyperedges of order $d$ are autocorrelated at lag $\tau$.

\begin{figure}[t!]
	\centering
		\includegraphics[width=\columnwidth]{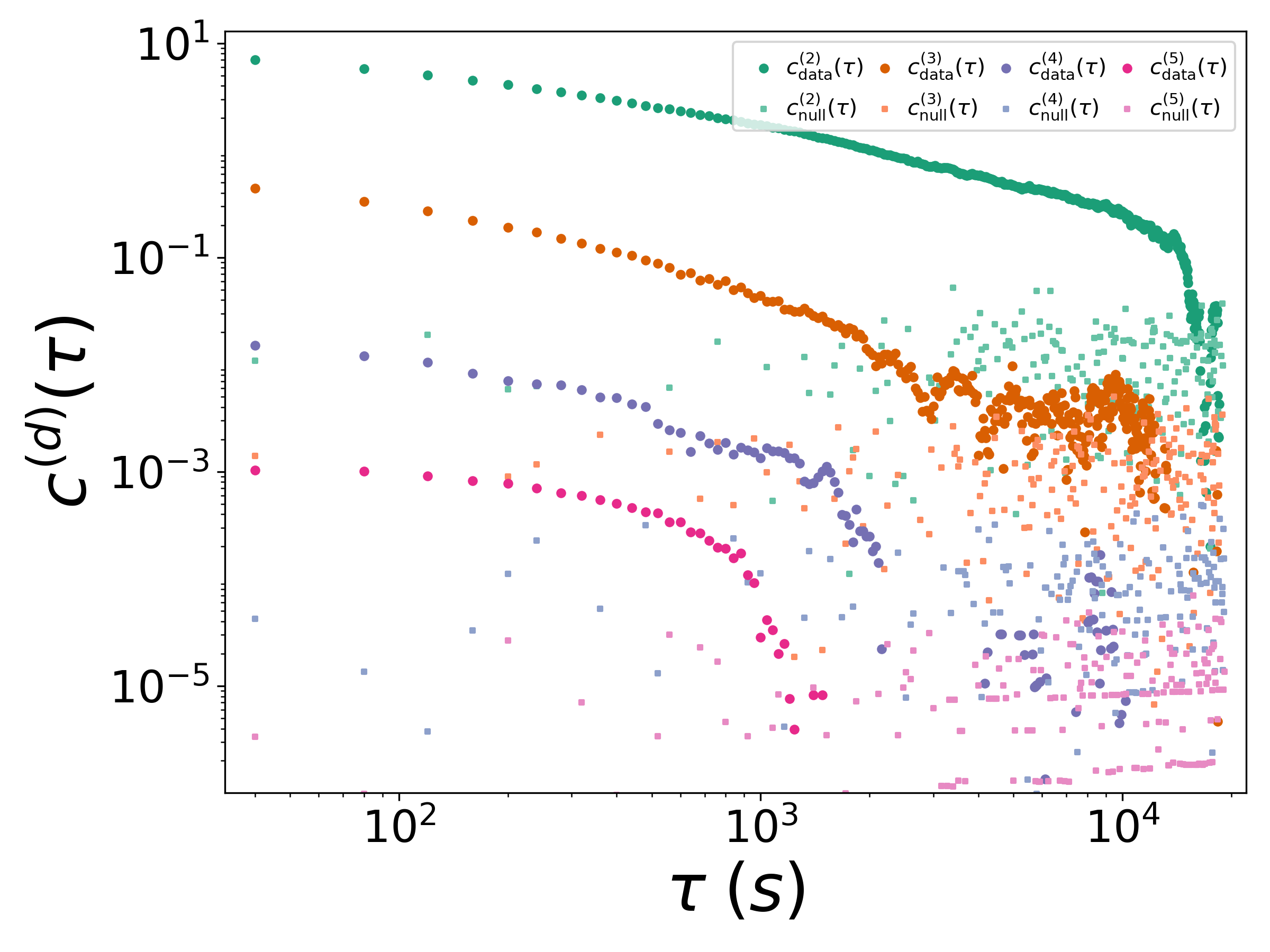}
		\caption{Intra-order correlations in human face-to-face interactions. Circles show the value of $c^{(d)}(\tau)$ for interactions in groups of different sizes, $d\in\{2,\dots,4\}$. Squares refer to a randomized null model where temporal correlations have been removed by reshuffling time-steps.}
	\label{fig:intra_order_correlations_conference}
\end{figure}

\begin{figure*}[t!]
	\centering
		\includegraphics[width=2\columnwidth]{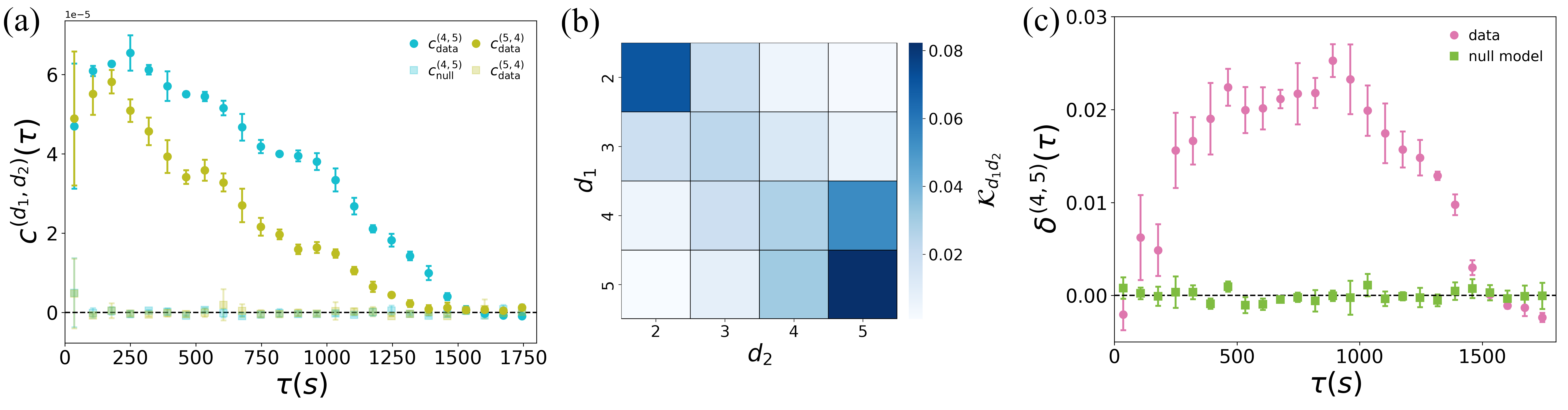}
		\caption{Cross-order correlations in human face-to-face interactions. (a) Cross-order correlation functions, $c^{(4,5)}(\tau)$ (cyan) and $c^{(5,4)}(\tau)$ (olive), describing the temporal dependencies between interactions of order four and interactions of order five. We compare the empirical system (circles) with a randomized null model with reshuffled time-steps (squares). (b) Normalized interaction matrix $\mathcal{K}_{d_1d_2}(\tau)$, encoding the temporal dependencies between any pairs of order $d_1,d_2\in\{2,\dots,4\}$ at time lag $\tau=600 s$. (c) Cross-order gap function $\delta^{(4,5)}(\tau)$ (purple circles), compared with the null model (green squares).}
	\label{fig:cross_order_correlations_conference}
\end{figure*}
Second, one can inquire whether interactions of two different orders $d_1$ and $d_2$ display a temporal interdependence, or otherwise evolve independently. 
To estimate such cross-order temporal correlations, we introduce the cross-order correlation matrix
\begin{equation}\label{eq:cross_order_correlation_matrix}
\begin{split}
\footnotesize
 \mathscr{C}^{(d_1,d_2)}(\tau) = 
\sum\limits_{t=1}^{T-\tau}\frac{\left[A^{(d_1)}(t)-\mu^{(d_1)}\right]\cdot\left[A^{(d_2)}(t+\tau)-\mu^{(d_2)}\right]^\intercal}{(T-\tau)(d_1-1)!(d_2-1)!},
\end{split}
\end{equation}
where $d_1,d_2\in\{2,\dots,D\}$. Note that, when $d_1=d_2=d$, we recover the intra-order correlation matrix $\mathscr{C}^{(d)}$.
We can then define a scalar cross-order correlation function as
\begin{equation}\label{eq:cross_order_correlation_function}
c^{(d_1,d_2)}(\tau) = \mathrm{tr}(\mathscr{C}^{(d_1,d_2)}(\tau)).   
\end{equation}
All the information about intra-order and cross-order correlations can be encoded in a
 $(D-1)\times (D-1)$ 
normalized interaction matrix $\mathcal{K}_{d_1d_2}(\tau) = c^{(d_1,d_2)}(\tau)/2\sqrt{\sigma^{(d_1}\sigma^{(d_2}}$, where $\sigma^{(d)}=c^{(d)}(0)$,
whose entry $\mathcal{K}_{d_1d_2}(\tau)$ describes how interactions of order $d_1$ at a given time are correlated with those of order $d_2$ occurring $\tau$ time steps later. 
Notice that matrix $\mathcal{K}_{d_1d_2}(\tau)$
is not symmetric, as the quantity $c^{(d_2,d_1)}(\tau)$ measures how order $d_1$ is correlated with order $d_2$ at $\tau$ time steps before, 
and is in general different from $c^{(d_1,d_2)}(\tau)$.  
The presence of a significant discrepancy between these two quantities captures asymmetries in the temporal dependencies between different orders of interaction. We quantify such an asymmetry in terms of a cross-order gap function
\begin{equation}\label{eq:cross_order_gap_function}
\delta^{(d_1,d_2)}(\tau) = \frac{c^{(d_1,d_2)}(\tau) - c^{(d_2,d_1)}(\tau)}{2\sqrt{\sigma^{(d_1)}\sigma^{(d_2)}}}.
\end{equation}
A positive value of $\delta^{(d_1,d_2)}(\tau)$ implies that the presence of groups of size $d_1$ correlates with the presence of groups of size $d_2$ after a time lag $\tau$, more than the other way around.

\vspace{3mm}
\noindent \emph{Temporal dependencies within and across orders emerge in social systems}. --- We now characterize temporal correlations in real-world systems with group interactions. 
We consider different social systems, for which we have high-resolution data about their temporal evolution. 
We here focus on a dataset describing face-to-face interactions over a period of 32h among the $N=403$ participants of a scientific conference \cite{cattuto2010dynamics,isella2011s}. 
Three further cases, namely the social interactions occurring in an office \cite{genois2015data}, in a hospital ward \cite{vanhems2013estimating} and in a university campus \cite{sapiezynski2019interaction} are described in the Supplemental Material (SM). 
We encode the fine-grained temporal information of the dataset in a temporal hypergraph $(\mathcal{V},\mathcal{H}(t))$, with $|\mathcal{V}|=403$ and $t\in\{1,\dots,T\}$. 
The set $\mathcal{H}(t)$ is constructed by assuming that $d$ individuals in contact at a given time $t$ interact together in a group of size $d$, thus corresponding to a hyperedge of order $d$ at time $t$.

We begin by studying how groups of a given size are temporally correlated. 
Fig.~\ref{fig:intra_order_correlations_conference} reports the intra-order correlation functions $c^{(d)}(\tau)$ for orders $d\in\{2,\dots,4\}$ (circles).  
Significant long-range temporal autocorrelations emerge for different orders of interaction, as indicated by the slow decays of $c^{(d)}(\tau)$ with $\tau$ in a double logarithmic scale, up to a threshold, which typically decreases with $d$. 
This indicates that groups of larger sizes generally remain autocorrelated for shorter times. 
Interestingly, we also observe a saturation effect for interactions in groups of size three, with a series of peaks revealing a weak periodicity at large timescales (see SM for further details).
Empirical results are compared with a null model (squares) obtained by reshuffling the sequence defining the temporal hypergraph. 

We now investigate whether interactions in groups of a given size $d_1$ can also be correlated to interactions in 
groups of size $d_2 \neq d_1$. The cross-order correlation functions $c^{(4,5)}(\tau)$ (cyan circles) and $c^{(5,4)}(\tau)$ (olive circles)  
for groups of sizes four and five, 
reported in Fig.~\ref{fig:cross_order_correlations_conference}(a),
clearly show the presence of significant 
cross-order correlations (see SM for an analysis of other group sizes). 
We observe that both functions display a peak followed by a slow decay, which can not be reproduced by the corresponding null model (squares). 
For clarity of presentation, we display a binned average of the cross-correlations functions, as well as the corresponding standard deviation. 
Fig.~\ref{fig:cross_order_correlations_conference}(b) shows the normalized interaction matrix $\mathcal{K}(\tau)$ at time lag $\tau=600 s$ (see SM for an analysis of different $\tau$). 
We observe a banded structure around the main diagonal, meaning that cross-order correlations are higher between groups of similar sizes. 
This indicates that, in the interactions at a scientific conference analyzed here, groups change gradually, with the loss or the addition of one or few members (see SM for the interaction matrix of different social systems, including the social contacts in a university campus where large groups reveal a more complex correlation pattern).
Finally, Fig.~\ref{fig:cross_order_correlations_conference}(c) shows the cross-order gap function $\delta^{(4,5)}(\tau)$ (purple circles). Values of $\delta^{(4,5)}(\tau) > 0$ in almost all the range of the time lag $\tau$ considered indicate that groups of size four at a given time are correlated to those of size five occurring $\tau$ time steps later, more than the other way around.
This result, which  again cannot be reproduced by the null model (green squares), suggests that the formation of a group of five individuals from a group of four is more probable than the loss of one member in groups of five individuals, indicating a preferred temporal direction in the dynamics of group formation/fragmentation of this social system (see SM for an analysis of $\delta^{(d_1,d_2)}(\tau)$ for other group sizes). 

\vspace{3mm}
\noindent  \emph{Temporal hypergraph models with higher-order memory}. --- To investigate the mechanisms shaping the onset of intra-order and cross-order correlation profiles, we introduce two models to generate temporal hypergraphs with higher-order memory, inspired by DAR processes~\cite{jacobs1978discrete,williams2019effects,williams2022shape}. 
The first model, named Discrete Auto Regressive Hypergraph (DARH) model, treats the binary states of each hyperedge $h^\alpha \in \{0,1\}$ (absent/present) as independent stochastic processes.  
Each hyperedge updates its state either drawing a state from its past, or randomly sampling a new state.  
With probability $q^{(d)}$, where $d\in\{2,\dots, D\}$, a hyperedge of order $d$ samples its state uniformly at random from its $m_s^{(d)}$ previous states. 
With probability $1-q^{(d)}$, the hyperedge state is drawn randomly following a Bernoulli process with probability $y^{(d)}$. 
Formally, the dynamics of a hyperedge of order $d$, $h^\alpha$, is given by $h^{\alpha}_{t} = Q_t h^{\alpha}_{t-\mu} + (1-Q_t)Y_t$, where $Q_t \sim \mathrm{Bernoulli}(q^{(d)})$, $Y_t \sim \mathrm{Bernoulli}(y^{(d)})$, and $\mu\sim\mathrm{Uniform}(1,m_s^{(d)})$.

Our first model displays intra-order, but no cross-order temporal correlations (see SM for a characterization of the DARH model). 
We then introduce a second model, the cross-memory DARH (cDARH) model, a variation of the DARH model where a hyperedge of order $d$ updates its state by drawing not only from its past but also from that of a hyperedge of a different order. 
We will refer to this latter property as cross-order memory. 
When copying from memory, each hyperedge draws from the past of other hyperedges with probability $p^{(d)}$, and from its own past with probability $1-p^{(d)}$. 
We assume hyperedges can copy from the memory of overlapping hyperedges only. 
This choice is motivated by previous studies on higher-order interactions in social networks that pointed out a tendency of groups to progressively add or remove members, one step at a time~\cite{cencetti2021temporal}. 
For illustration, let us consider groups of size two and three: a $2$-hyperedge $\{i,j\}$ selects one of the $(N-2)$ possible $3$-hyperedges containing nodes $i$ and $j$ and draws from its previous $m_c^{(3,2)}$ states. 
Similarly, a $3$-hyperedge $\{i,j,k\}$ selects one of the three $2$-hyperedges that can be formed from it, i.e., $\{i,j\}$, $\{j,k\}$, and $\{i,k\}$, and copies a state from its previous $m_c^{(2,3)}$ steps. 
Such a mechanism can be straightforwardly extended to hyperedges of other orders. 
Formally, we can write the dynamics of a hyperedge of order $d$ as
\begin{equation}
    h^{\alpha}_{t} = Q_t h^{\varepsilon_t(\alpha)}_{t-\mu} + (1-Q_t)Y_t.
\end{equation}
As for the DARH model, we have that $Q_t \sim \mathrm{Bernoulli}(q^{(d)})$ and $Y_t \sim \mathrm{Bernoulli}(y^{(d)})$, while the new random variable $\varepsilon_t(\alpha)$ selects the hyperedge from which the past is copied. 
Consequently, the variable $\mu$ does not only depend on the order of $h^\alpha$ but also on the value of $\varepsilon_t(\alpha)$. 
For instance, in the case of $2$-hyperedges, $\mu\sim\mathrm{Uniform}(1,m_s^{(2)})$ if $\varepsilon_t(\alpha)=\alpha$, while $\mu\sim\mathrm{Uniform}(1,m_c^{(3,2)})$ if $\varepsilon_t(\alpha)\neq\alpha$, when copying from $3$-hyperedges.

\begin{figure}[t!]
	\centering
		\includegraphics[width=1\columnwidth]{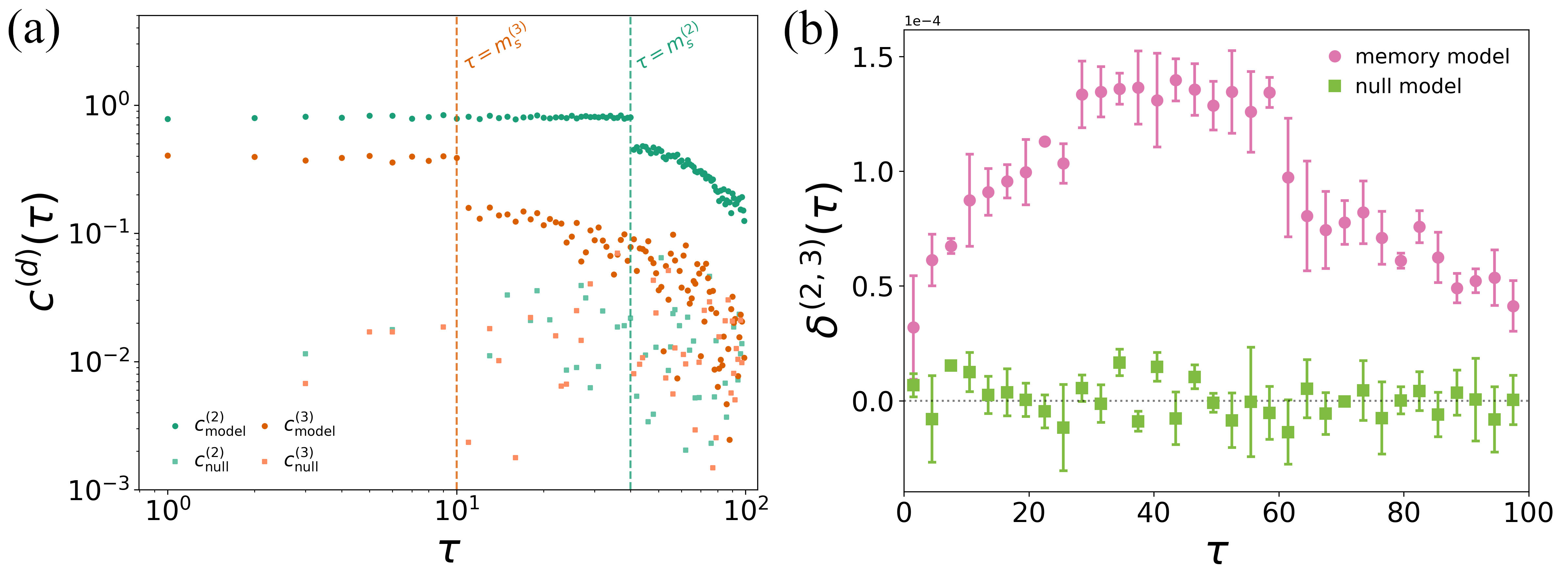}
		\caption{Temporal correlations in the cDARH model. (a) Intra-order correlations $c^{(d)}$, with $d\in\{2,3\}$, for hyperedges of order two (green circles and squares) and three (orange circles and squares). The dashed vertical lines correspond to the value of the intra-order memory of hyperedges of order two (green) and three (orange), respectively. (b) Cross-order gap function $\delta^{(2,3)}$ between hyperedges of order two and three. We compare intra-order and cross-order correlations for the hypergraph generated using the cDARH model with a randomized null model.}
	\label{fig:cDARH_intra_cross_correlations}
\end{figure}

We generate with the cDARH model temporal hypergraphs with $N=10$ nodes, maximum hyperedge order $D=3$ and a temporal range of $T=3\cdot 10^4$ time steps.  
Fig.~\ref{fig:cDARH_intra_cross_correlations} reports the results obtained by setting $p^{(2)}=0$ and $p^{(3)}=0.6$, meaning that hyperedges of order three can copy from the past of hyperedges of order two, while hyperedges of order two evolve independently. 
We also set $m_s^{(2,3)}=60$, for the cross-order memory of $3$-hyperedges, $m_s^{(2)}=40$ and $m_s^{(3)}=10$ for the intra-order memories of hyperedges of order two and three, respectively.  
Fig.~\ref{fig:cDARH_intra_cross_correlations}(a) reveals the presence in hyperedges of order two (green circles) and three (orange circles) of significant intra-order temporal correlations. 
In particular, we notice that the functions $c^{(2)}$ and $c^{(3)}$ remain constant for $\tau\leq m^{(d)}$, while they decay exponentially after that value, with the same rate of decay (see SM for a deeper analysis). 
Such a minimal model reveals that memory can be the driving mechanism for the emergence of intra-order temporal correlations, with different orders possessing different degrees of memory, explaining the hierarchical structure of correlation observed in the data. 
The profiles of $c^{(d)}(\tau)$ do not match exactly those of empirical data, which are characterized by a power-law decay. 
This is not surprising, as in the cDARH model a single set of values $m_s^{(2)}$ and $m_s^{(3)}$ is considered, while real-world social interactions can be shaped by different scales of memory~\cite{williams2022shape}. 
Fig.~\ref{fig:cDARH_intra_cross_correlations}(b) shows that $\delta^{(2,3)}(\tau)>0$ for different values of $\tau$ (purple circles), meaning that hyperedges of order two are correlated to hyperedges of order three occurring later in time more than the other way around.
We observe a striking similarity between this trend and that observed in Fig.~\ref{fig:cross_order_correlations_conference}(c) for the empirical data. 
This result indicates that cross-order memory is a fundamental factor for the emergence of cross-order gaps in real-world social interactions. 
The two peaks for $\delta^{(4,5)}(\tau)$ observed in the empirical system suggest again a more complex dependence on memory, possibly due to multiple temporal scales. 

\noindent \emph{Discussion}.--- 
In this Letter, we have introduced a framework to characterize different dimensions of memory in networked systems with higher-order interactions. 
We have shown that real-world social systems display long-range temporal correlations at different group sizes, organized in a hierarchy across multiple scales. 
Moreover, we found that group interactions are characterized by cross-order correlations, suggesting the emergence of non-trivial interactions between mesoscopic memories, overall leading to a complex memory structure. 
%
In the context of social systems, such cross-order interactions can be interpreted in terms of the schisming phenomenon \cite{palla2007quantifying,stehle2010dynamical,egbert1997schisming,mastrangeli2010roundtable}, where group sizes in human interactions, e.g. conversations, fluctuate, nucleate, and display complex dynamics. 
The formalism presented here can be naturally extended to other higher-order complex systems traditionally modeled in terms of networks of interactions, such as the human brain and biological ecosystems. 
In conclusion, our work sheds light on the unexpectedly complex and multifaceted nature of memory that emerges in real-world interacting systems and we hope can open new avenues also in the higher-order dynamics of emergent coherent structures in a variety of physical systems, from multifragmentation in nuclear physics to vortex interaction in the atmosphere or other fluid dynamical systems.

\noindent  \emph{Acknowledgments}.--- L.G. acknowledges C. Zappalà for the insightful discussions and comments.
L.G. and F.B. acknowledge support from the Air Force Office of Scientific Research under award number FA8655-22-1-7025. L.L. acknowledges funding from the Spanish Research Agency (AEI) via projects DYNDEEP (EUR2021-122007), MISLAND (PID2020-114324GB-C22), and the María de Maeztu project CEX2021-001164-M.

\bibliography{biblio}

\pagebreak
\widetext
\begin{center}
\textbf{\large Supplemental Material of the manuscript \\
Higher-order correlations reveal complex memory in temporal hypergraphs}
\end{center}
\setcounter{equation}{0}
\renewcommand{\theequation}{S\arabic{equation}}
\setcounter{figure}{0}
\renewcommand{\thefigure}{S\arabic{figure}}

\section{Intra-order correlations in social systems with higher-order interactions}
In the main text, we focused on a dataset describing face-to-face interactions in a scientific conference. In this section, we present the analysis of the intra-order temporal correlations for three other social systems. Two of the datasets come from the Sociopatterns project, and they describe face-to-face interactions \textit{(i)} in an office, among $N=92$ workers, over 11 days \cite{genois2015data}; and \textit{(ii)} in a hospital ward, among $N=75$ patients, doctors, nurses, and administrative staff, over 72h \cite{vanhems2013estimating}. The third dataset contains instead proximity data of $N=692$ students at the campus of the Technical University of Denmark (DTU) in Copenhagen, for a period of one month \cite{sapiezynski2019interaction}. 

\begin{figure}[h!]
	\centering
		\includegraphics[width=\columnwidth]{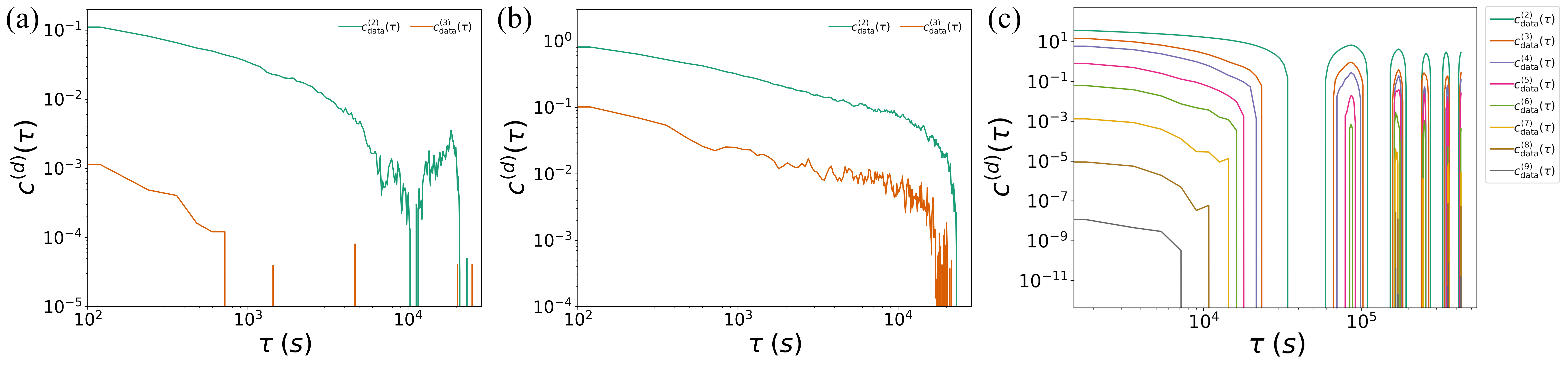}
		\caption{Intra-order correlation functions in human face-to-face interactions in the office (a), in the hospital ward (b), and in the university campus (c), for different group sizes.}
	\label{fig:intra_order_social_systems}
\end{figure}

In this section, we analyze temporal correlations within groups of the same size. Fig.~\ref{fig:intra_order_social_systems} depicts the values of the intra-order correlation functions for several group sizes $d$, for the three empirical systems examined, namely face-to-face interactions in the workplace (a), the hospital ward (b), and in the university campus (c). For each system, we consider only those group sizes for which we have a sufficient statistics. In general, for each social system, a hierarchy of long-range, slowly-decaying temporal autocorrelations emerges across different group sizes. Also, we note that intra-order correlations are lost abruptly after a typical time threshold that decreases with the group size $d$, i.e., larger groups remain  temporally correlated for shorter time. 

We observe periodicity patterns in the interactions occurring in the workplace and in the university campus. For this latter in particular, we note a series of peaks at large timescales that show the same hierarchical structure observed at small timescales. See Fig.~\ref{fig:periodic_intra_cross_correlations} for a minimal model of a temporal hypergraph with periodicity.

\section{Cross-order correlations show nontrivial temporal organization of higher-order interactions}
In this section, we deepen our analysis on temporal correlations between groups of different sizes. In the main text, we showed that statistically significant cross-order correlations emerge. In particular, we focused on the cross-order correlations between groups of sizes four and five in the face-to-face interactions in the scientific conference. We here show that other pairs of group sizes in the same system are temporally correlated.

\begin{figure}[h!]
	\centering
		\includegraphics[width=0.5\columnwidth]{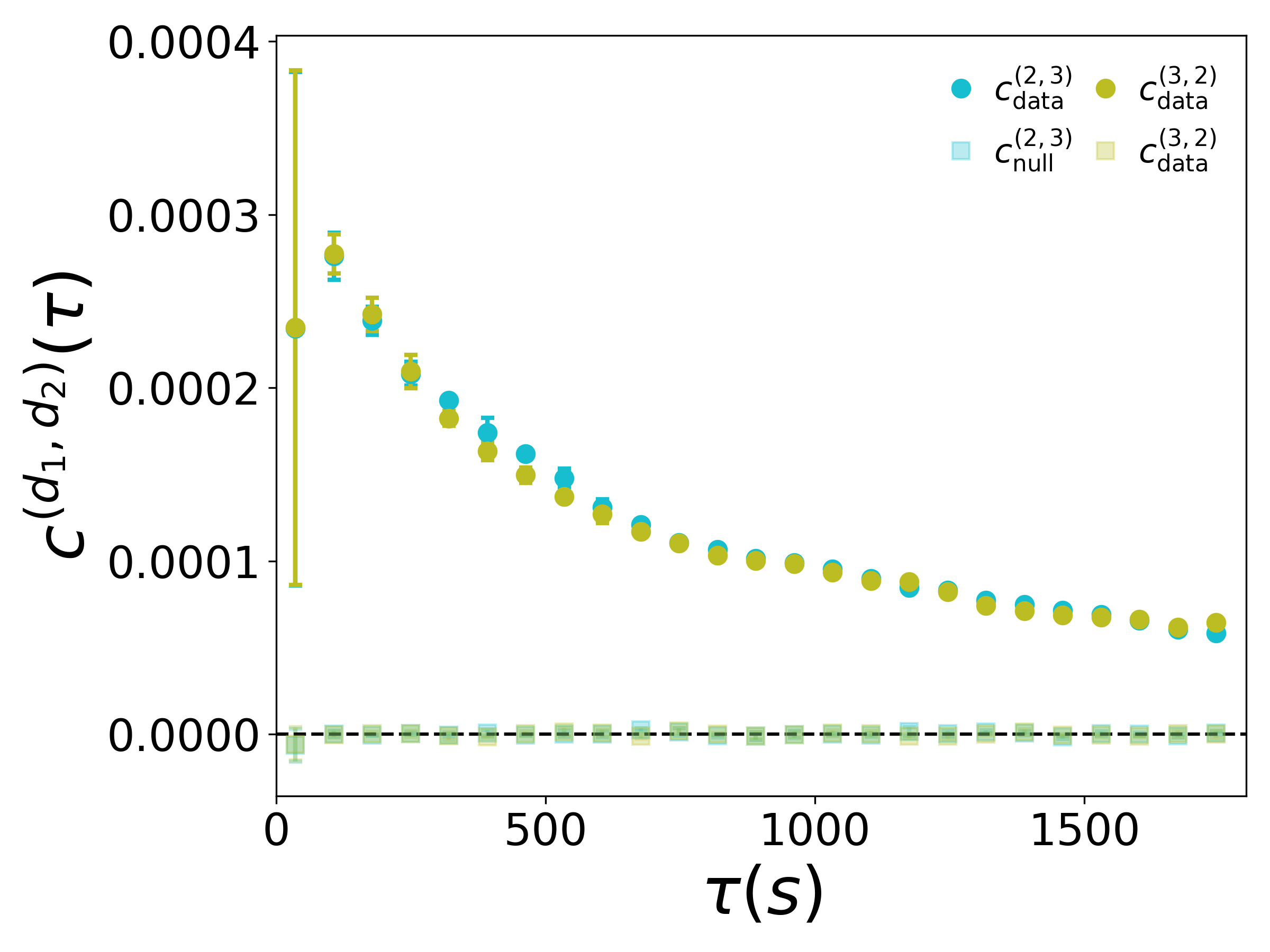}
		\caption{Cross-order correlation functions for groups of sizes two and three, i.e., $c^{(2,3)}(\tau)$ (cyan) and $c^{(3,2)}(\tau)$ (olive), in the face-to-face interactions in the scientific conference. The empirical system (dark circles) is compared with a randomized null model with reshuffled time-steps (light squares).}
	\label{fig:cross_correlation_12_conf}
\end{figure}
Fig.~\ref{fig:cross_correlation_12_conf} displays the cross-order correlation functions $c^{(2,3)}(\tau)$ (cyan) and $c^{(3,2)}(\tau)$ (olive), describing cross-order
correlations between groups of sizes two and three. We also report the cross-order correlation functions for a randomized null model where correlations have been removed by time-reshuffling (lighter squares)
Both functions exhibit an exponential decay that does not match the trends of the cross-order correlation functions $c^{(4,5)}(\tau)$ and $c^{(5,4)}(\tau)$ shown in Fig.~2(a) of the main text, suggesting that different pairs of group sizes can show different kinds of interdependencies. 

\begin{figure}[h!]
	\centering
		\includegraphics[width=\columnwidth]{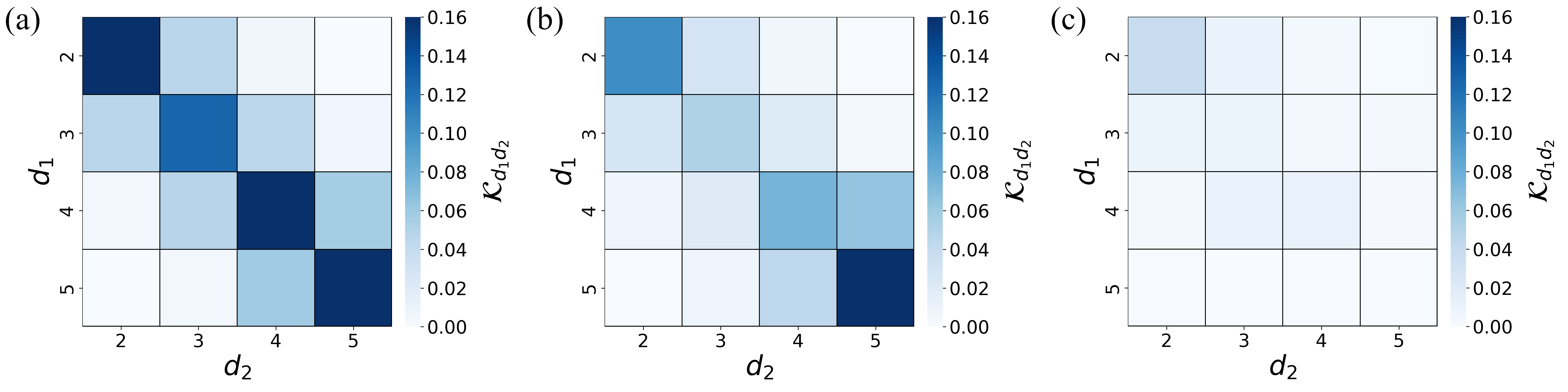}
		\caption{Normalized cross-order interaction matrix $\mathcal{K}(\tau)$ at time lags $\tau=60s$ (a), $\tau=300s$ (b), and $\tau=1500s$ (c), in the face-to-face interactions in the scientific conference.}
	\label{fig:cross_correlation_matrix_conf}
\end{figure}
Fig.~\ref{fig:cross_correlation_matrix_conf} displays the normalized interaction matrix $\mathcal{K}(\tau)$ at different time lags, namely $\tau=60s$ (a), $\tau=300s$ (b), and $\tau=1500s$ (c). We observe that cross-order temporal correlations can emerge between different group sizes. In particular, as discussed in the main text, the matrix $\mathcal{K}(\tau)$ shows a banded structure around the main diagonal, meaning that correlations between groups of similar sizes is higher. The banded structure is lost at large time lags, as correlations vanish at larger timescales.

\begin{figure}[h!]
	\centering
		\includegraphics[width=0.5\columnwidth]{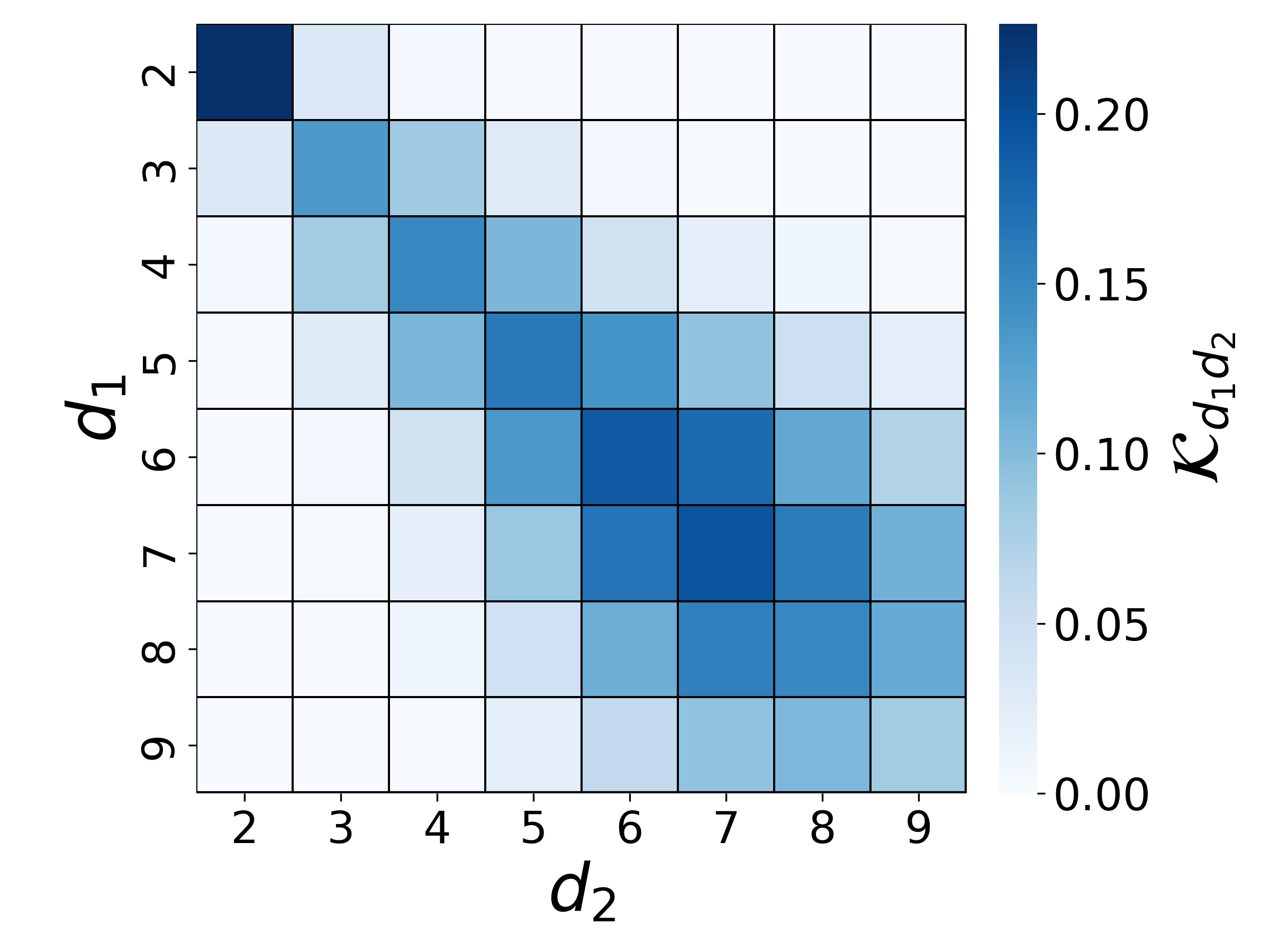}
		\caption{Normalized interaction matrix $\mathcal{K}(\tau)$ at time lag $\tau=300s$, in the DTU campus in Copenhagen.}
	\label{fig:cross_correlation_matrix_copenhagen}
\end{figure}
The banded structure of the matrix $\mathcal{K}(\tau)$ suggests that, in the scientific conference dataset, groups tend to change gradually, with the loss or the addition of one member. 
We now want to investigate whether groups of larger size display the same property. 
To answer this question, we focus on the interactions among students occurring in the DTU campus in Copenhagen, in which larger groups have a sufficient statistics.
Fig.~\ref{fig:cross_correlation_matrix_copenhagen} displays the normalized interaction matrix $\mathcal{K}(\tau)$ at time lag $\tau=300s$. 
Once again, the matrix $\mathcal{K}(\tau)$ reveals a characteristic banded structure. 
However, compared to the scientific conference case, we note that groups of larger size show a different pattern. 
Indeed, we can observe significant cross-order correlations between groups of dissimilar sizes, e.g., groups of size nine are temporally correlated with groups of size five. 
This result suggests that larger groups evolve differently compared to smaller ones. 
In particular, while small-size groups change gradually, losing or gaining one member at a time, large-size groups show more complex dynamics, as they can also split in smaller groups or emerge from them.  

\begin{figure}[h!]
	\centering
		\includegraphics[width=0.66\columnwidth]{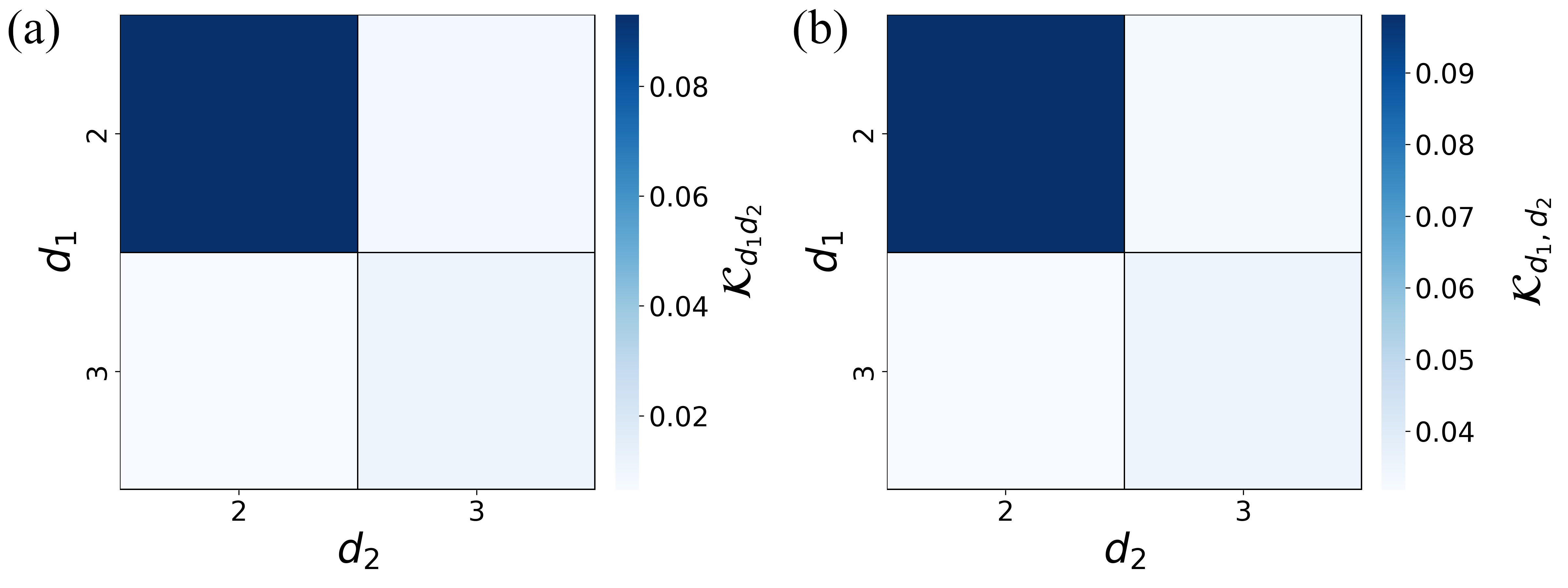}
		\caption{Normalized interaction matrix $\mathcal{K}(\tau)$ at time lag $\tau=300s$ for the face-to-face interactions occurring in the office (a) and in the ospital ward (b).}
	\label{fig:cross_correlation_matrix_work_hosp}
\end{figure}
For completeness, in Fig.~\ref{fig:cross_correlation_matrix_work_hosp} we report the normalized interaction matrices evaluated at time lag $\tau=300s$ relative to the interactions in groups of size two and three occurring in the office (a) and in the hospital ward (b). 

\begin{figure}[h!]
	\centering
		\includegraphics[width=0.5\columnwidth]{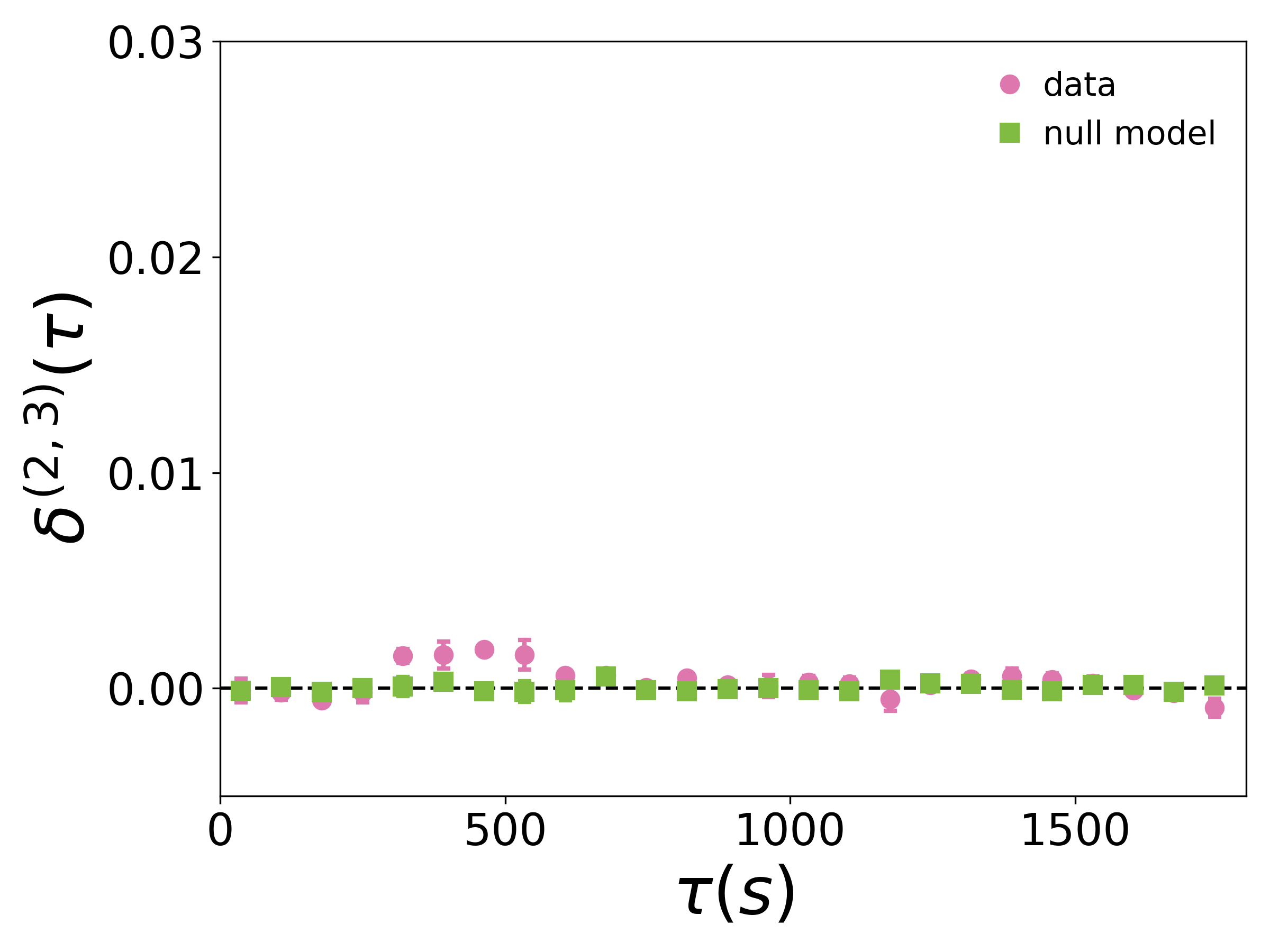}
		\caption{Cross-order gap function $\delta^{(2,3)}(\tau)$, for interactions in groups of size two and three in the scientific conference (purple circles), compared with the null model (green squares).}
	\label{fig:cross_gap_12_conf}
\end{figure}
In the main text, we showed that the cross-order correlation functions relative to groups of sizes $d_1$ and $d_2$ can differ significantly, as groups of a given size can anticipate those of another one, while the vice versa might not be true, or the magnitude of the two effects might be distinct.
This discrepancy between the two functions is quantified in terms of the cross-order gap function $\delta^{(d_1,d_2)}(\tau)$. 
To complement the analysis provided in the main text, here we consider the interactions in the scientific conference, studying the possible gaps in the cross-order correlation functions for other pairs of group sizes. 
Fig.~\ref{fig:cross_gap_12_conf} shows the cross-order gap function $\delta^{(2,3)}(\tau)$ for the real social network (purple circles). 
Note that the figure axis scales are the same as those of Fig.~2(c) of the main text, for a proper comparison. 
We observe that $\delta^{(2,3)}(\tau) $ does not significantly differ from the null model where temporal snapshots are reshuffled. 
This indicates that neither the groups of size two nor those of size three tend to anticipate the other, suggesting that groups of two people gain a member with the same probability as groups of three people lose one.  

\begin{figure}[h!]
	\centering
		\includegraphics[width=\columnwidth]{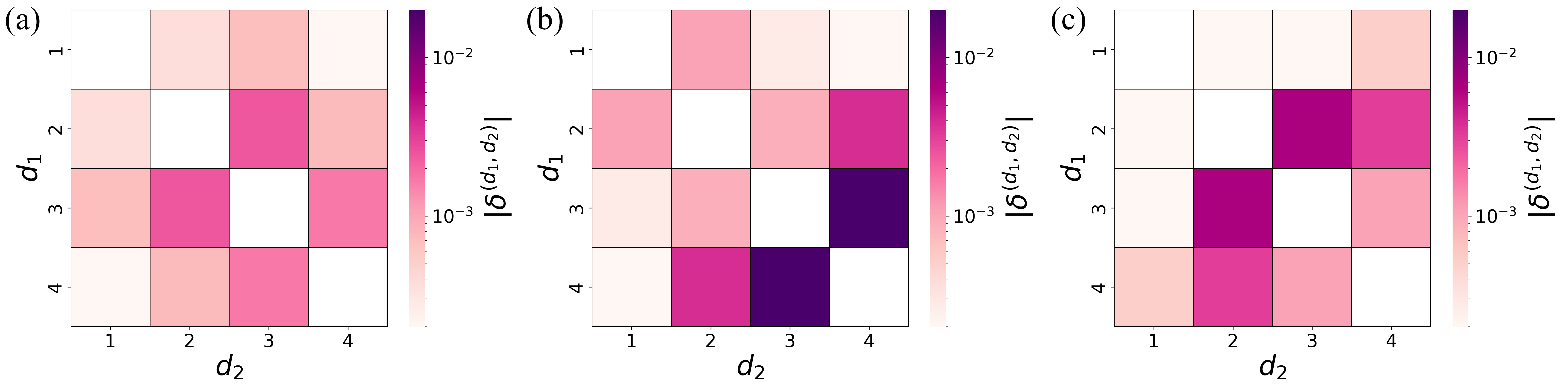}
		\caption{Value of $|\delta^{(d_1,d_2)}|$ for different group sizes $d_1$ and $d_2$, at different times lags, namely $\tau=60s$ (a), $\tau=600s$ (b), and $\tau=1500s$ (c), in the face-to-face interactions in the scientific conference.}
	\label{fig:delta_matrix_conf}
\end{figure}
Fig.~\ref{fig:delta_matrix_conf} shows the value of $|\delta^{(d_1,d_2)}|$ for different values of $d_1$ and $d_2$, at different times lags, namely $\tau=60s$ (a), $\tau=600s$ (b), and $\tau=1500s$ (c). 
We observe that different other pairs $d_1$, $d_2$ display a non-zero value of $|\delta^{(d_1,d_2)}|$, again suggesting a non-trivial temporal organization between different orders of interactions.

\section{Analysis of the DARH model}
In this section we analyze the Discrete Auto-Regressive Hypergraph (DARH) model, namely a minimal model of temporal higher-order network with memory. 
As described in the main text, in the DARH model, each hyperedge $h^\alpha$ evolves independently and at each time step $t$ it updates its state either by drawing from its memory or randomly.
The draw from the past occurs with probability $q^{(d)}$, with $d\in\{2,\dots, D\}$ for hyperedges of sizes two up to $D$, while with probability $1-q^{(d)}$, the next state is chosen at random, according to a Bernoulli process with probability $y^{(d)}$. 
When the state of the hyperedge is drawn from the past, we assume hyperedges of size $d$ to copy one of the previous $m_s^{(d)}$ states. 
Formally, the dynamics of a hyperedge of order $d$, $h^\alpha$, is given by 
\begin{equation}
h^{\alpha}_{t+1} = Q_t h^{\alpha}_{t-\mu} + (1-Q_t)Y_t,
\end{equation}
where $Q_t \sim \mathrm{Bernoulli}(q^{(d)})$, $Y_t \sim \mathrm{Bernoulli}(y^{(d)})$, and $\mu\sim\mathrm{Uniform}(1,m_s^{(d)})$.

\begin{figure}[h!]
	\centering
		\includegraphics[width=\columnwidth]{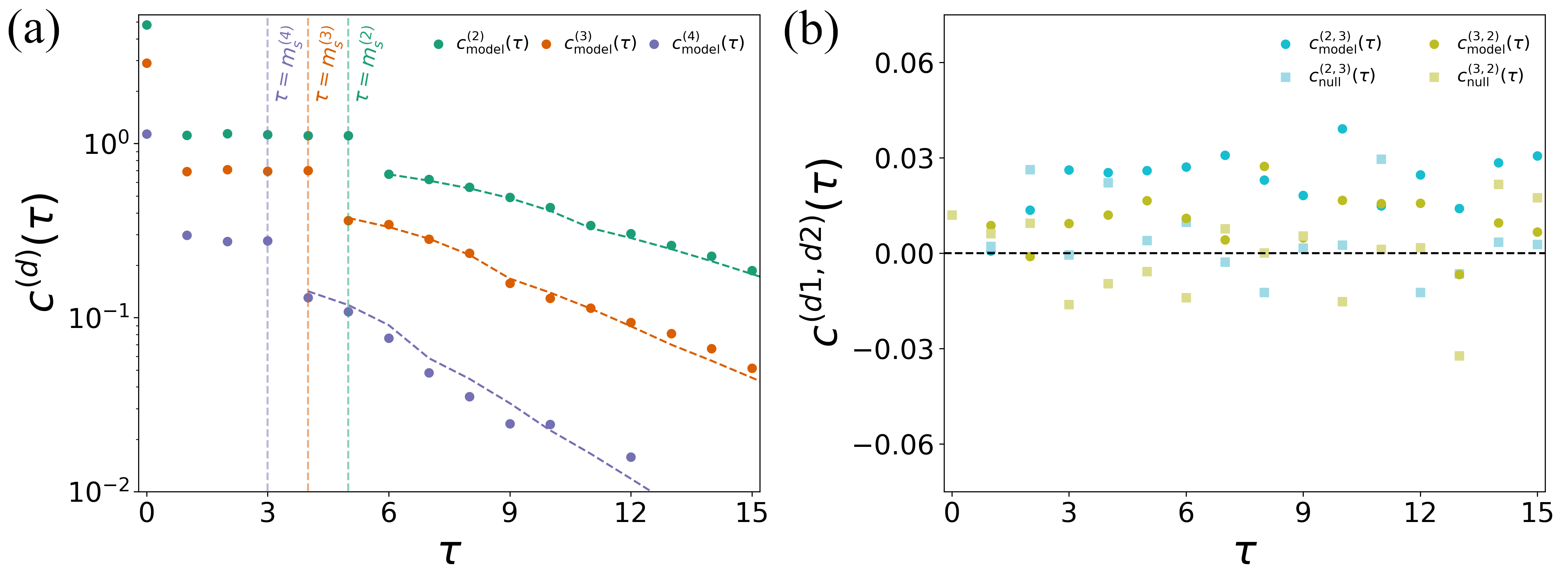}
		\caption{Intra-order correlation functions and cross-order gap functions in the DARH model. (a) Intra-order correlations. Each color represents a different order of interaction. Vertical dashed lines represent the values $\tau=m_s^{(d)}$, with $d\in{1,2,3}$. Other colored dashed lines display the theoretical prediction of the Yule-Walker equations. (b) Cross-order correlation functions between hyperedges of orders two and three, namely $c^{(2,3)}(\tau)$ (cyan) and $c^{(3,2)}(\tau)$ (olive). Dark dots represent the values obtained for the DARH model, while light squares show the correlation functions for a time-reshuffled hypergraph.}
	\label{fig:DARH_intra_cross_correlations}
\end{figure}

We generate a temporal hypergraph using the DARH model. 
We build a higher-order network of $N=5$ nodes, composed by $T=10^5$ snapshots. 
We consider hyperedges up to order four. 
The memory parameters for the interactions of orders two, three and four are set to $m_s^{(2)}=5$, $m_s^{(3)}=4$, and $m_s^{(4)}=3$, respectively. 
The probabilities of drawing from the past for the different orders of interaction are set to $q^{(2)}=0.60$, $q^{(3)}=0.55$, and $q^{(4)}=0.50$. 
Finally, the probabilities to activate each hyperedge in the memory-less process are set to $y^{(2)}=0.4$, $y^{(3)}=0.05$, and $y^{(4)}=0.02$. 
Fig.~\ref{fig:DARH_intra_cross_correlations} shows the intra-order correlation functions for each order of interaction (a) and the cross-order correlation functions $c^{(2,3)}(\tau)$ and $c^{(3,2)}(\tau)$ (b) for the temporal hypergraph generated with the DARH model. 
For simplicity, for the cross-order correlation functions we limited the analysis to hyperedges of orders two and three. 

In Fig.~\ref{fig:DARH_intra_cross_correlations}(a), we observe that for all hyperedge orders the intra-order correlation remain constant for $1 \leq \tau \leq m_s^{(d)}$ (vertical dashed lines), while it decays exponentially after that value. 
At $\tau=0$ the value of the autocorrelation functions is trivially higher, and depends on the average number of hyperedges of a given order over the whole temporal range.

For each order $d$, the intra-order correlation function can be reproduced by the Yule-Walker equations (dashed lines):
\begin{equation}\label{eq:yule_walker}
    c^{(d)}(\tau) = \frac{q^{(d)}}{m_s^{(d)}}\sum\limits_{k=1}^{m_s^{(d)}}c^{(d)}(\tau-k),
\end{equation}
which determine the autocorrelation function of a DAR($m_s^{(d)}$) process. 
In particular, one can prove \cite{williams2019effects} that $c^{(d)}(\tau)$ is constant for $\tau \leq m_s^{(d)}$, with its value given by 
\begin{equation}
    c^{(d)}(\tau \leq m_s^{(d)}) = c^{(d)}(0)\left[m_s^{(d)}\left(\frac{1}{q^{(d)}}-1\right)+1\right]^{-1},
\end{equation}
which allows us to solve Eqs.~\eqref{eq:yule_walker} up to any finite lag $\tau$. 
The DARH model highlights that memory can be a driving factor for the emergence of intra-order temporal correlations. 
The model reveals that the hierarchical structure of the correlation ranges observed in the data (see both main text and previous sections of this Supplemental Material) is due to the different degrees of memory possessed by the hyperedges. 

Fig.~\ref{fig:DARH_intra_cross_correlations}(b) shows instead that hyperedges of sizes two and three do not display cross-order correlation. 
Indeed, both $c^{(2,3)}(\tau)$ (cyan) and $c^{(3,2)}(\tau)$ (olive) are close to zero (dark dots), and do not differ significantly from the values obtained for a null model where the time-steps of the temporal hypergraph have been reshuffled (light squares). 
Indeed, in the DARH model each hyperedge evolves independently from other hyperedges, either those having the same size and those of different sizes. 
As a consquence, we observe no cross-order correlations among hyperedges of different sizes.

\section{Exponential decay in the cDARH model}
In this section, we analyze intra-order correlations in the cDARH model. 
In the main text, we have considered a model configuration where the maximum order of the hyperedges is $D=3$, where hyperedges of order two can copy from their own past, while hyperedges of order three can also copy from the past of the overlapping hyperedges of order two (see the main text for further details). 
In this setting, we find that the intra-order correlation function $c^{(2)}(\tau)$ can be reproduced by the Yule-Walker equations \eqref{eq:yule_walker}, while $c^{(3)}(\tau)$ can not. 
This is not surprising, as hyperedges of order two evolve independently, so their state follows a independent DAR process (see previous section for more detail), while the evolution of hyperedges of order three also depends on the dynamics of three overlapping hyperedges of order two.

\begin{figure}[h!]
	\centering
		\includegraphics[width=0.5\columnwidth]{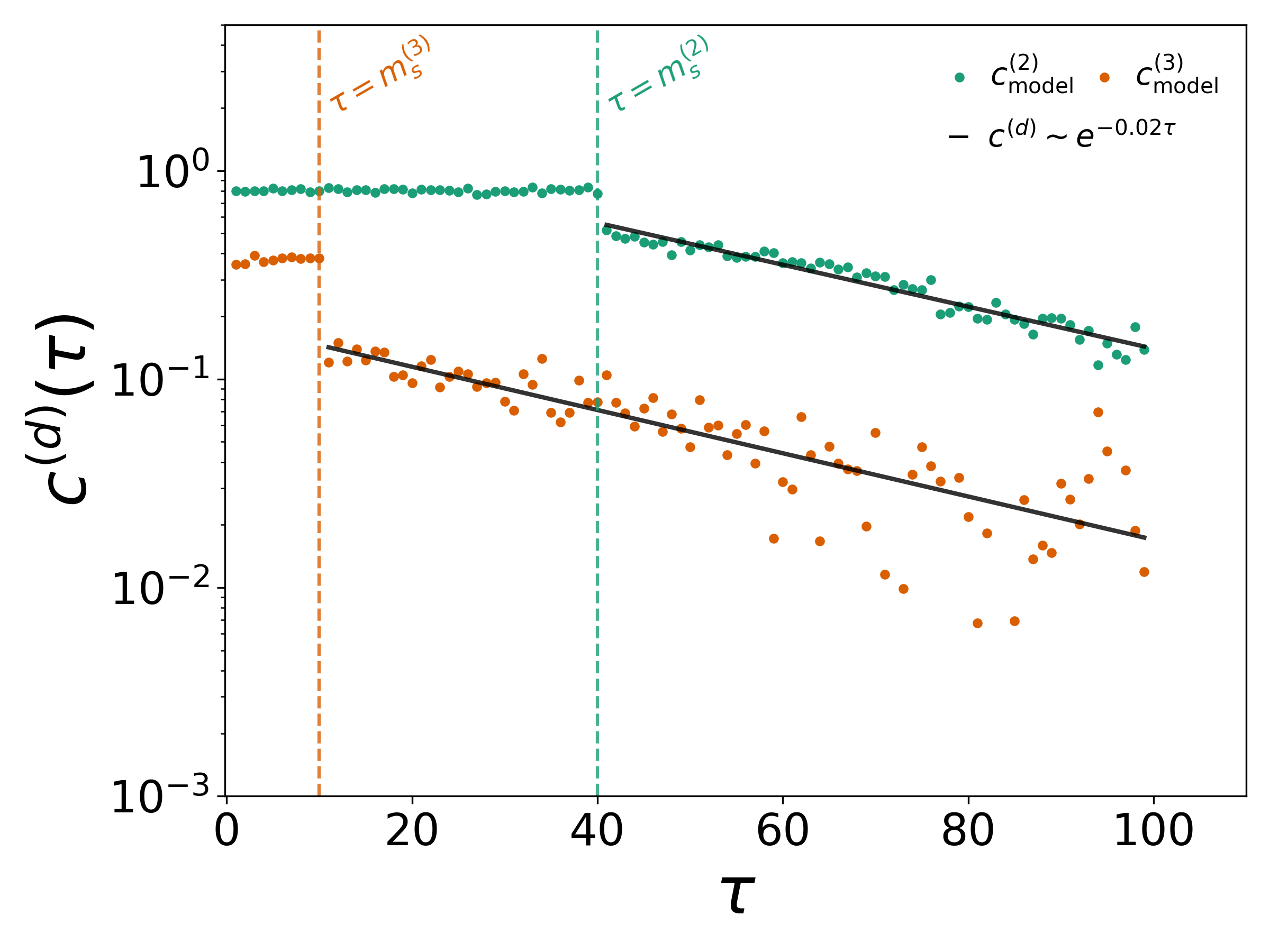}
		\caption{Intra-order correlations in the cDARH model. Green and orange circles represent the value of $c^{(d)}$ for hyperedges of orders two and three, respectively. The dashed vertical lines correspond to the value of the intra-order memory of hyperedges of order two (green) and three (orange). The solid black line displays the best fit of the exponential decay for both $c^{(2)}$ and $c^{(3)}$.}
	\label{fig:cDARH_intra_correlations}
\end{figure}
To gain insights about the behavior of the cDARH model, we fit the exponential decay of both $c^{(2)}(\tau)$ and $c^{(3)}(\tau)$. 
We find that both exponential decays have the same rate, i.e., $c^{(d)} \sim e^{\beta\tau}$, with $\beta=0.02$, as reported in Fig.~\ref{fig:cDARH_intra_correlations}. 
This result is essentially due to the fact that hyperedges of order three have memory of the overlapping hyperedges of order two.
This mechanism might also explain why interactions in real-world systems are characterized by long-range intra-order correlation functions that decays with similar rates (see Fig.~\ref{fig:intra_order_social_systems} and Fig.~1 in the main text). 
Note that, to better appreciate the rate comparison, here we consider a lin-log scale, whereas in Fig.~3(a) of the main text we used a log-log scale.

\section{Correlation analysis of a periodic temporal hypergraph}
Social systems show a variety of temporal patterns, from periodicity to decaying autocorrelations. 
In the main text, we concentrated on this latter feature, discussing a minimal model of temporal higher-order networks with memory. 
In this section, we focus instead on temporal periodic patterns in systems characterized by higher-order interactions. 
To do so, we build a temporal hypergraph of $N$ nodes where each order of interaction has different periodicity. 
For simplicity and with no loss of generality, we limit our analysis to groups of size two and three. 
First, we construct a sequence of length $T_2$ of random $2$-uniform hypergraphs, namely hypergraphs where all hyperedges have order two. 
Each hyperedge has a probability $p_2$ to be created. 
Each hypergraph of the sequence is nothing more than a Erd\"os-R\'enyi random network, i.e., $\mathrm{ER}(N,p_2)$. 
We then construct another sequence, having length $T_3$, and composed by random $3$-uniform hypergraphs, i.e., all hyperedges connect exactly three nodes. 
A hyperedge has a probability $p_3$ to be created. 
For each order of interaction, we concatenate several of the corresponding sequences one after another, thus building two temporal uniform hypergraphs. 
We repeat the sequences so that both temporal hypergraphs have $T$ time-steps each. 
Finally, for each time-step we join the two corresponding static uniform hypergraphs. 
By doing this, we now have a single temporal hypergraph of length $T$ where the $2$-hyperedges are repeated every $T_2$ time-steps, while the $3$-hyperedges have instead a period $T_3$.

\begin{figure}[h!]
	\centering
		\includegraphics[width=\columnwidth]{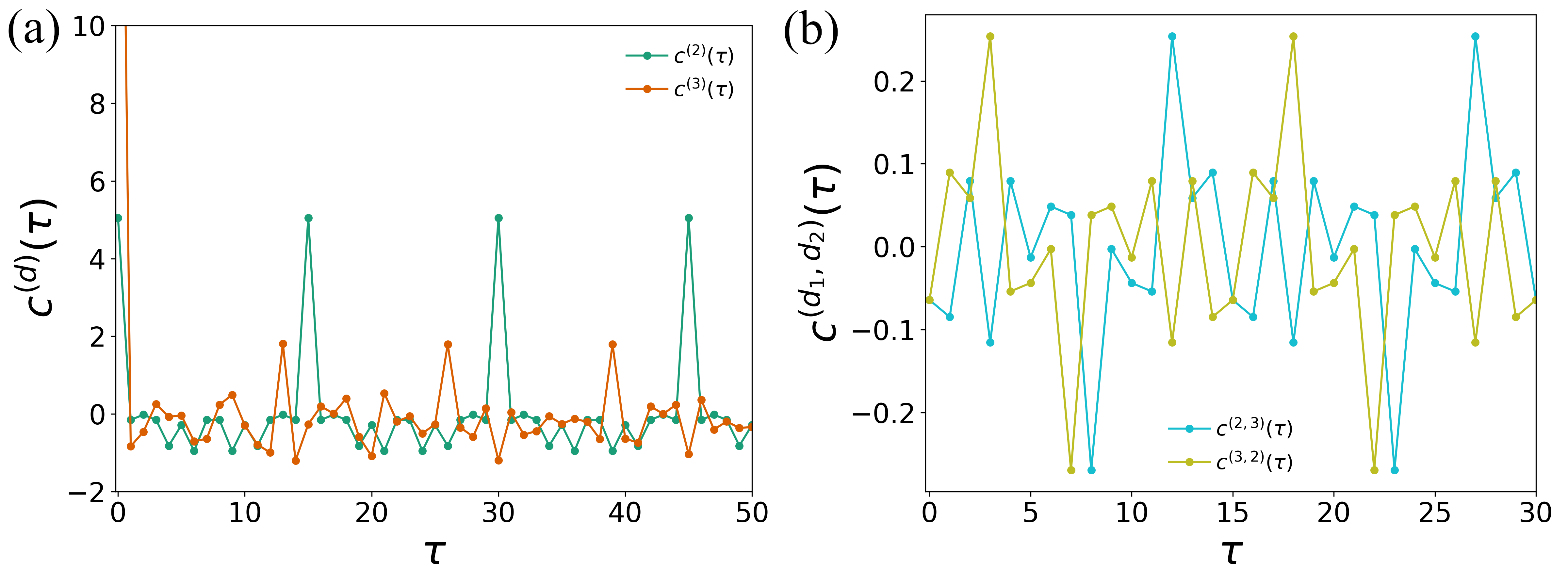}
		\caption{Intra-order correlation functions and cross-order gap functions in a minimal model of periodic temporal hypergraphs. (a) Intra-order correlations in hyperedges of order two (green) and three (orange), respectively. (b) Cross-order correlation functions between hyperedges of orders two and three, i.e., $c^{(2,3)}(\tau)$ (cyan) and $c^{(3,2)}(\tau)$ (olive).}
	\label{fig:periodic_intra_cross_correlations}
\end{figure}
Fig.~\ref{fig:periodic_intra_cross_correlations} shows the intra-order correlation functions (a) and the cross-order correlation functions (b) for a periodic temporal hypergraph of $N=10$ nodes.
The hypergraph is composed by $T=1.95\cdot10^5$ snapshots. 
Each $2$-hyperedge has a probability $p_2=0.1$ of existing and it repeats with period $T_2=15$. 
The $3$-hyperedges have instead a probability $p_3=0.03$ of being active and they have period $T_3=8$. 

Both orders of interactions clearly show a periodic behavior.
Indeed, we observe that the intra-order correlation functions peak at $\tau=k T_2$ and $\tau= k T_3$, with $k\in \mathbb{N}$, for two-body and three-body interactions, respectively, while $c^{(2)}(\tau)$ and $c^{(3)}(\tau)$ remain close to zero for other values of $\tau$. 
The periodicity of the temporal hypergraph is also reflected in the cross-order correlation functions $c^{(2,3)}(\tau)$ (cyan) and $c^{(3,2)}(\tau)$ (olive). 
In particular, we observe that they peak at different harmonics and that in general $c^{(2,3)}(\tau) \neq c^{(3,2)}(\tau)$, as the periodic oscillations of the two orders of interaction are not in phase.

\end{document}